# AM-DCGAN: Analog Memristive Hardware Accelerator for Deep Convolutional Generative Adversarial Networks

Olga Krestinskaya[1], Bhaskar Choubey[2], Alex Pappachen James[1]
[1]Nazarbayev University, Astana;  [2]Siegen University, Germany

## Introduction

Generative Adversarial Network (GAN) is a well known computationally complex algorithm requiring signficiant computational resources in software implementations[1] including large amount of data to be trained. This makes its implementation in edge devices with conventional microprocessor hardware a slow and difficult task. In this paper, we propose to accelerate the computationally intensive GAN using memristive neural networks in analog domain. We present a fully analog hardware design of Deep Convolutional GAN (DCGAN) [2] based on CMOS-memristive convolutional and deconvolutional networks simulated using 180nm CMOS technology.

## Analog CMOS-memristive DCGAN architecture

GAN utilises two neural networks with competitive learning approach, one of generator of new data and another of discriminator of this new data with the training data., . The proposed analog hardware implementation of DCGAN is illustrated in Fig. 1 (a). The Generator Network has been implemented as a  Memristive Deep Deconvolutional Neural Network (Deep DCNN), while the Discriminator Network is a Memristive Deep Convolutional Neural Network (Deep CNN). During  training, the Generator produces the fake images from input noise while theDiscriminator is trained to identify whether the image is fake or real using supervised training method. Discriminator acts as a binary classifierproducing a single output,. Deconvolution and Convolution operations in Generator and Discriminator Circuits are performed using memristive crossbar based dot product multiplication. CMOS activation functions and other system components are designed to process signals in analog domain. The analog error propagation and weight update unit updates the memristive weights in Generator and Discriminator. A full circuit level implementation of Generator and Discriminator Networks is shown in Fig. 2 and Fig. 3 in Supplementary Material.

## Results

The system was  tested for MNIST handwritten digits generation using CNN and DCNN with 3 layers (2 convolutional layers with 128 and 64 filters of size 5x5 and a dense layer in both). A set of example generated images after 100 training epochs of DCGAN is presented in Fig. 1 (b). Circuit components of both architectures was analyzed using SPICE. Fig. 4 and Fig. 5 (Supplementary Material) illustrate the transient response for 100ns pulses and outputs of designed activation functions, respectively.  Small CNN and DCNN systems with 2 convolutional layers and 2 filters of the size of 3x3, and single dense layer network showed that total on-chip area and power consumption of such circuit are approximatelly 0.188mm$^2$ and 7W, respectively, with details shown in Table 1. The system has been designed in 0.18um CMOS technology for demonstration. Implementation of the proposed design in smaller geometries will further reduce on-chip area and power consumption significantly.

The performance of Memristive GAN can be effected by device variability, restricted number of resisitve levels and non-idealities of memristive devices. We performed variabilty analysis in memristive crossbar with Fig. 7 illustrating the effect memristor variabilities on the output error in different blocks of the architecture. Fig. 6 demonstrates the effect of memristor variabilities on the leakage currents. We also demonstrate the effect of memristor variabilities (Fig. 7) and restricted number of resistive levels of memristors (Fig. 8) on the quality of generated images. The simulation results show that the architecture can tolerate up to 30% of variability in memrsitive devices and 64 resistive levels without significantly degrading the quality of generated output. Fig. 9 illustrates the effect of number of training epochs on the generated image showing that  50 epochs (corresponding to maximum 3M update cycles of memristive devices)can produce desirable image quality. Based on this, power consumption for memristor update process, training time for different length of training pulses and different ways of parallel and sequential update shown in Table II is calculated.

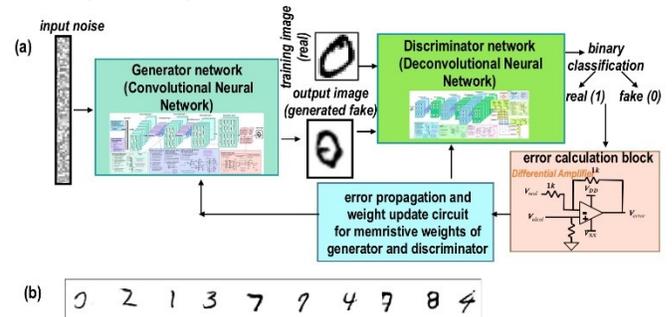

*Fig. 1 (a) Analog hardware implementation of DCGAN with on-chip learning, and (b) example of generated images*.

## Conclusions

By treating the memristive neural networks as hardware subroutines, it is possible to develop an analog system that can speed up complex neural computing for near sensor data processing of GAN. The main advantages of analog DCGAN are small on-chip area and possibility to implement it on edge devices and integrate directly to analog sensors to process information in real time and speed up the training.

# Supplementary Material

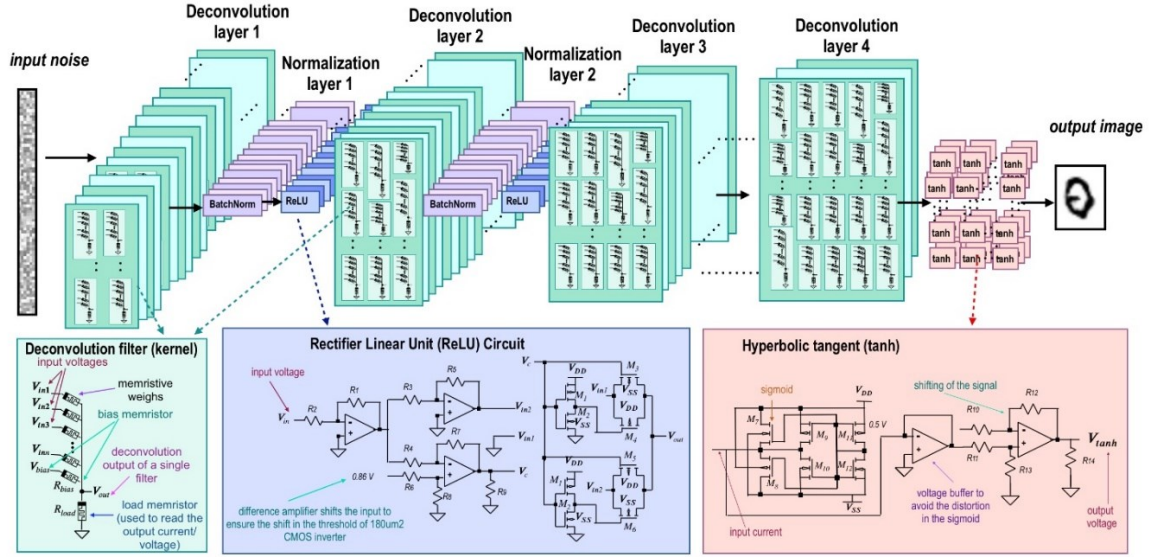

Fig.2 GAN Generator based on Memristive Deconvolutional Neural Network architecture. The deconvolution is performed using memristive crossbar based dot product multiplication. The output of each crossbar is read from the load memristor., This ensures that the output voltage does not exceed the maximum filter input. The deconvolution filter outputs are normalized using batch normalization circuit and ReLU. To reduce on-chip area and power consumption, batch normalization circuit can be removed, as ReLU function in analog domain has an ability to clip the maximum voltage level and does not allow the output to exceed $V_{DD}$. The deconvolution is performed several times to achieve the desired size of the image. The output image from DCGAN is generated from the final deconvolutional filtering layer using hyperbolic tangent activation function.

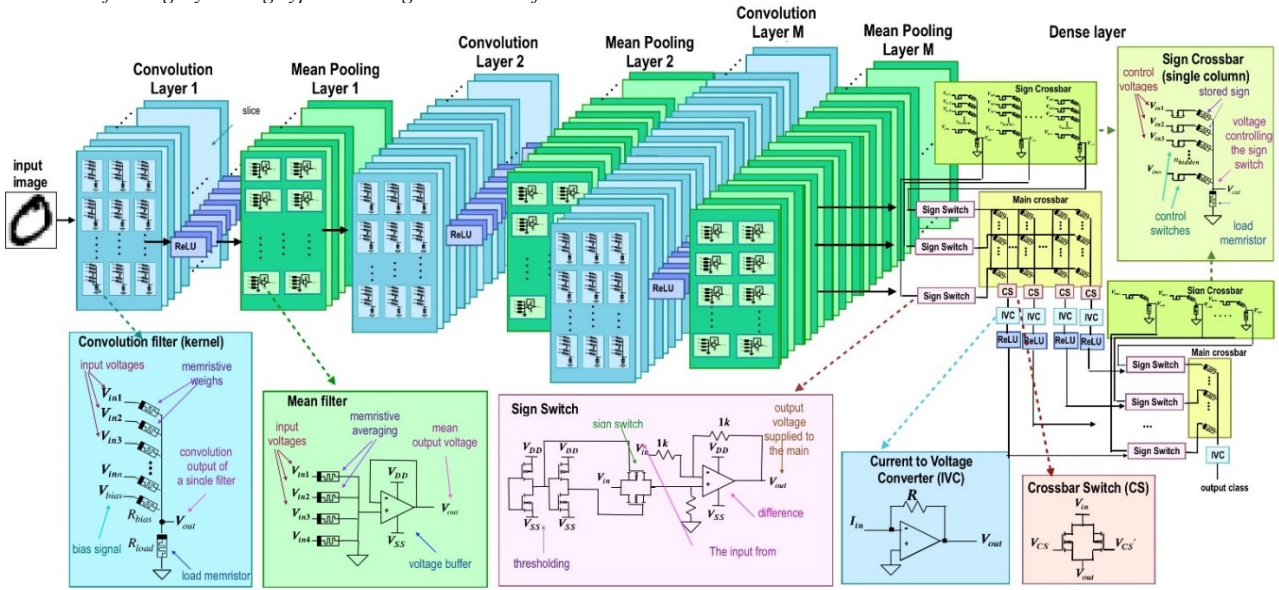

Fig. 3 GAN Discriminator based on Convolutional Neural Network architecture. The architecture consists of several memristive convolutional and mean pooling layers and dense network. The convolutional layer output is normalized by ReLU followed by the mean pooling operation performed by memristive averaging circuit. After the last convolutional layer, the output image is processed by dense network, which is a conventional ANN with linear activation function. The sign of memristive weights in the dense layer is controlled by the sign crossbar and sign switch circuit, based on the thresholding circuit, which inverts the input to the dense network when required. The columns in the dense network crossbar are processed sequentially. The output current from the column is converted to voltage by OpAmp based Current-to-Voltage Converter (IVC) and processed with ReLU circuit.

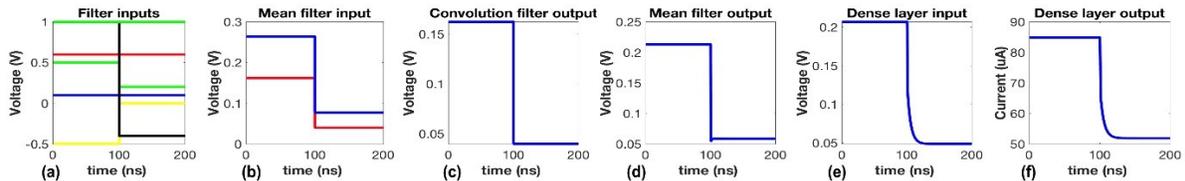

Fig. 4 Transient analysis for Discriminator (CNN): (a) inputs to one of the convolutional filters, (b) corresponding output of the convolutional filter, (c) input to mean filtering, (d) output of the mean filter, (e) corresponding input to the dense layer crossbar, (f) output current form the dense layer column.

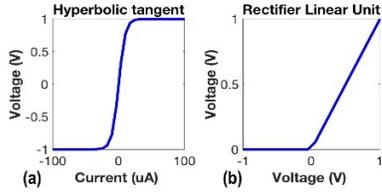

Fig. 5 Outputs of the activation functions: (a) hyperbolic tangent and (b) ReLU with switch.

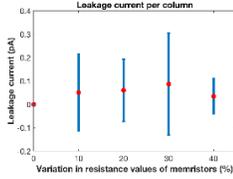

Fig. 6. Leakage current calculation in memristive crossbars for different noise levels.

TABLE I
APPROXIMATE POWER CONSUMPTION AND ON-CHIP AREA OF CMOS COMPONENTS OF THE ARCHITECTURE.

| Circuit Component | Power Consumption (mW) | On-chip area (µm²) |
|---|---|---|
| Dropout Switch | 0.0033 | 14.5 |
| OpAmp | 7.4000 | 558.3 |
| Thresholding Circuit | 0.0586 | 0.8 |
| ReLU | 23.300 | 951.1 |
| Crossbar Switch | 5.0000 | 5.0 |

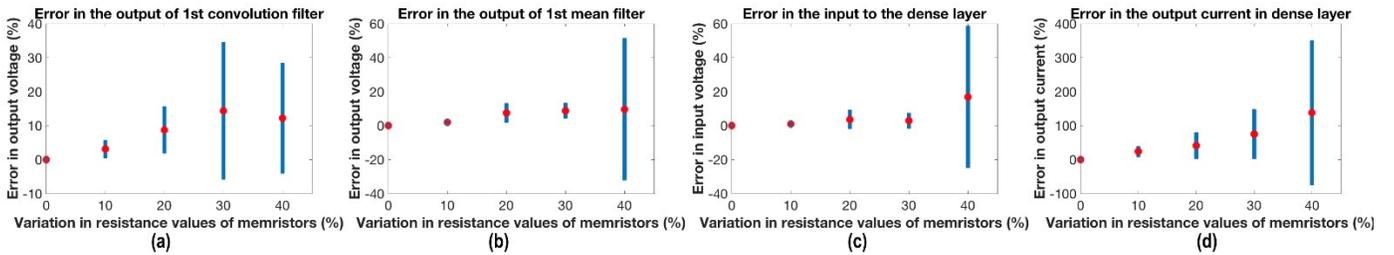

Fig. 7 Error caused by the variabilities in memristive weights in CNN: (a) error in the output of the convolutional filter, error in the mean filter, (c) error in the dense layer input after, and (d) error in the output current in the dense layer.

Fig. 7 Effect of variabilities in memristive devices on the quality of generated images for memristors with 128 stable resistive levels.

Fig. 8 Effect of the restricted number of resistive levels on generated image quality.

Fig. 9 Effect of the number of training iterations (memristor updates) of the number quality of generated images.

TABLE II
TRAINING TIME AND POWER FOR WO₂ MEMRISTIVE DEVICES WITH $R_{ON}$=4K AND $R_{OFF}$=25K, THRESHOLD VOLTAGE 0.8V, WRITE VOLTAGE 1V AND DIFFERENCE WRITE TIMES WITH 1.7 MILLION MEMRISTIVE WEIGHTS TO UPDATE FOR 50 EPOCHS FOR 60,000 MNIST IMAGES (ARCHITECTURE WITH 3 CROSSBAR LAYERS IN GENERATOR AND DISCRIMINATOR, MAXIMUM 128 CONVOLUTIONAL FILTERS OF THE SIZE OF 5X5 AND IMAGES OF THE SIZE OF 28X28)

| Update Configuration | Power Consumption (W) | | Training time (s) | | |
|---|---|---|---|---|---|
| | maximum | minimum | 10ns* | 100ns* | 1000ns* |
| Parallel update of CNN and DCNN with sequential update of independent layers | 0.00150 | 0.00024 | 19e5 | 19e6 | 19e7 |
| Parallel update of independent columns in layers and independent layers | 0.48800 | 0.07808 | 23.552 | 235.52 | 2355.20 |
| Update of memristors in 4 cycles of independent layers in series (rows and columns) | 17.2000 | 2.80000 | 0.72 | 7.20 | 72.0 |
| Parallel update of memristors using 2 cycles of independent layers in series | 35.4000 | 5.60000 | 0.36 | 3.60 | 36.0 |

*Write time for memristor